\documentstyle[prd,preprint,aps]{revtex}
\begin{document}
\draft

\title{Non-trivial classical backgrounds with vanishing quantum
corrections} 
\author{L.~Sriramkumar$^{1}$\thanks{E-Mail:~lsk@iucaa.ernet.in}, 
R.~Mukund$^{2}$\thanks{E-Mail:~mukund@niharika.phy.iitb.ernet.in} 
and T.~Padmanabhan$^{1}$\thanks{E-Mail:~paddy@iucaa.ernet.in}} 
\address{$^{1}$IUCAA, Post Bag 4, Ganeshkhind, Pune 411 007, INDIA.} 
\address{$^{2}$Department of Physics, IIT, Powai, Mumbai 400 071, INDIA.}
\maketitle

\begin{abstract}
Vacuum polarization and particle production effects in classical 
electromagnetic and gravitational backgrounds can be studied by 
the effective lagrangian method.
Background field configurations for which the effective lagrangian 
is zero are special in the sense that the lowest order 
quantum corrections vanishes for such configurations. 
We propose here the conjecture that there will be neither particle 
production nor vacuum polarization in classical field 
configurations for which all the scalar invariants are zero.
We verify this conjecture, by explicitly evaluating the effective 
lagrangian, for {\it non-trivial}\/ electromagnetic and gravitational 
backgrounds with vanishing scalar invariants. 
The implications of this result are discussed.
\end{abstract}
\vskip 16pt
\newpage

\section{Introduction}

The effective lagrangian approach is probably the most unambiguous 
approach available at present to study the evolution of quantum fields 
in classical backgrounds~\cite{schwinger51,dewitt75,bd75,parker79}.
In this approach, an effective lagrangian is obtained for the classical
background field by integrating out the degrees of freedom corresponding 
to the quantum field. 
The effective lagrangian thus obtained, in general, has a real and an
imaginary part. 
The real part is interpreted as the `vacuum-to-vacuum' transition
amplitude, {\it i.e.}~the amplitude for the quantum field to remain in 
the initial vacuum state at late times (vacuum polarization) and the 
existence of a nonzero imaginary part is attributed to the instability 
of the vacuum (particle production). 

Several non-perturbative features of the theory can be understood 
if the effective lagrangian can be evaluated {\it exactly}\/ for an 
{\it arbitrary}\/ background field configuration.
But, the evaluation of the effective lagrangian for an arbitrary 
classical background proves to be an uphill task. 
Therefore, there has been numerous attempts in
literature\cite{bi70,cr75,dc76,fhh79,hh79,n84,s88,s89,p91} 
to evaluate the effective lagrangian explicitly for a 
{\it given}\/ electromagnetic or gravitational background.

Symmetry considerations suggest that it should be possible to express 
the effective lagrangian, at least formally, in terms of invariant scalars 
describing the classical background (gauge invariant quantities involving 
the field tensor $F_{\mu\nu}$ and its derivatives in the case of
electromagnetism and coordinate invariant scalars involving the 
Riemann curvature tensor $R_{\mu\nu\lambda\rho}$ and its derivatives
in the case of gravity). 
The existence of an imaginary part to the effective lagrangian---and other 
features---should be related to the actual values of some of these scalars.
Consider, for example, the simple case a quantized complex scalar field 
interacting with an electromagnetic background that is constant both in 
space and time. 
Schwinger, in his classic paper~\cite{schwinger51}, had evaluated the 
effective lagrangian for such a background by integrating out the 
degrees of freedom corresponding to the quantum field. 
(Schwinger had in fact considered the quantum field to be a spinor field, 
but his result also holds good for the complex scalar field we consider 
in this paper.)   
He showed that the resulting expression for the effective lagrangian depends 
only on the two gauge invariant quantities ${\cal G}=F^{\mu\nu}F_{\mu\nu}$ 
and ${\cal F}=\epsilon^{\mu\nu\lambda\rho} F_{\mu\nu}F_{\lambda\rho}$. 
Further, the effective lagrangian had an imaginary part only if ${\cal G}<0$, 
thereby implying that constant magnetic fields cannot produce particles 
while constant electric fields can. 
Schwinger's result, of course, had been obtained only for constant
$F_{\mu\nu}$'s and it is not easy to evaluate the effective lagrangian 
for a more general case (see~\cite{sriram96} for an attempt in this 
direction). 
Also, for an arbitrary electromagnetic background, there is no a priori 
reason as to why the effective lagrangian cannot depend on invariant 
quantities involving the derivatives of $F_{\mu\nu}$'s, for instance, 
say, $\partial_{\lambda} F^{\mu\nu}\partial^{\lambda}F_{\mu\nu}$.

The situation is still worse in the case of gravitational backgrounds. 
The gravitational analogue of Schwinger's electromagnetic example would 
be the case of a constant gravitational field, {\it i.e.} a spacetime
whose $R_{\mu\nu\lambda\rho}$'s are constants.
It would certainly be a worthwhile effort to evaluate the effective 
lagrangian for such a background.
Though, considerable amount work has been done in this direction in 
literature (see, for instance, references~\cite{cr77,bp79,hc84,srini96}), 
we are yet to have a covariant criterion for particle production by 
constant gravitational fields (analogous to the criterion ${\cal G}<0$ 
Schwinger had obtained for the constant electromagnetic background).
Also, since the gravitational interaction is not renormalizable, it is not 
easy at all to regularize the effective lagrangian (see, for {\it e.g.}, 
sections $6.11$ and $6.12$ of reference~\cite{dewitt75} in this context).

In this paper, we investigate a related but more restricted question.
We ask: Can one find non-trivial background field configurations for 
which the (regularized) effective lagrangian vanishes identically? 
That is, we are interested in finding classical field configurations 
in which neither vacuum polarization nor particle production takes place.
Such configurations certainly enjoy some special status because these
are the ones for which lowest order semiclassical corrections vanish.
What kind of classical field configurations will have this feature?

The effective lagrangian for the {\it constant} electromagnetic background
reduces to zero when the gauge invariant quantities ${\cal F}$ and 
${\cal G}$ are set to zero.
Apart form this case, at least one more non-trivial electromagnetic field 
configuration is already known in literature for which the effective 
lagrangian proves to be zero.
Schwinger, in his pioneering paper~\cite{schwinger51} also calculates 
the effective lagrangian for a plane electromagnetic wave background 
(for which gauge invariant quantities ${\cal F}$ and ${\cal G}$ are 
zero) and shows that it vanishes identically. 
These results suggest the following conjecture. 
{\it The effective lagrangian will be zero if all the scalar 
invariants describing the background vanish identically.}
In this paper, we present examples of non-trivial electromagnetic and
gravitational backgrounds with vanishing scalar invariants to support 
our conjecture. 
We evaluate the effective lagrangian explicitly using Schwinger's proper 
time formalism for the case of a quantized complex scalar field and show 
that it identically vanishes in these backgrounds.

This paper is organized as follows.
In section~\ref{sec:em} we present an example from electromagnetism 
and in section~\ref{sec:grav} we present an example from gravity. 
We explicitly evaluate the effective lagrangian and show that it 
vanishes identically in these backgrounds. 
Finally, in section~\ref{sec:conclns}, we discuss the wider implications 
of our analysis.

\section{Effective lagrangian for the electromagnetic example}\label{sec:em}
\subsection{Preliminaries}

The system we shall consider in this section consists of a complex scalar 
field $\Phi$ interacting with an electromagnetic field represented by the 
vector potential $A^{\mu}$. 
It is described by the action
\begin{eqnarray} 
{\cal S}[A^{\mu}, \Phi]=\int d^4x\, {\cal L}(A^{\mu}, \Phi)
=\int d^4x\, \biggl\{
\left(\partial_{\mu}\Phi+iqA_{\mu}\Phi\right) 
&&\left(\partial^{\mu} \Phi^* - iq A^{\mu}\Phi^*\right)\nonumber\\
& &-m^2\Phi\Phi^* - {1 \over 4} F^{\mu\nu}F_{\mu\nu}\biggl\},
\label{eqn:emact}
\end{eqnarray}
where $q$ and $m$ are the charge and the mass associated with a single 
quantum of the complex scalar field, the asterisk denotes complex 
conjugation and 
\begin{equation}
F_{\mu\nu} = \partial_{\mu}A_{\nu} - \partial_{\nu} A_{\mu}.
\end{equation}
The electromagnetic field is assumed to behave classically, hence $A_{\mu}$ 
is just a $c$-number while the complex scalar  field is assumed to be a 
quantum field so that $\Phi$ is an operator valued distribution.
In such a situation, we can obtain an effective lagrangian for the 
classical electromagnetic background by integrating out the degrees 
of freedom corresponding to the quantum field as follows
\begin{equation}
\exp \,i\int d^4x\, {\cal L}_{eff}(A^{\mu})
\equiv\int{\cal D}\Phi\, \int{\cal D}\Phi^*\,
\exp \,i\, {\cal S}[A^{\mu}, \Phi],\label{eqn:leffem}
\end{equation}
where we have set $\hbar=c=1$ for convenience.
The effective lagrangian can then be expressed as
${\cal L}_{eff}= {\cal L}_{em}+{\cal L}_{corr}$,
where ${\cal L}_{em}$ is the lagrangian density for the free 
electromagnetic field, {\it viz.} the third term under the 
integral in action~(\ref{eqn:emact}), and ${\cal L}_{corr}$ is given by
\begin{eqnarray}
\exp \,i\int d^4x\, {\cal L}_{corr}(A^{\mu})
=\int{\cal D}\Phi\, \int {\cal D}\Phi^*\exp \,i\int d^4x\,
\biggl\{\left(\partial_{\mu}\Phi+iqA_{\mu}\Phi\right) 
&&\left(\partial^{\mu} \Phi^* - iq A^{\mu}\Phi^*\right)\nonumber\\
&& - m^2\Phi\Phi^*\biggl\}.
\end{eqnarray}
Integrating the action for the scalar field in the above equation by parts 
and dropping the resulting surface terms, we obtain that~\cite{ryder85} 
\begin{equation}
\exp\,i\int d^4x\, {\cal L}_{corr}(A^{\mu})= 
\int{\cal D}\Phi\,\int{\cal D}\Phi^*\, 
\exp -i\int d^4x\, \Phi^*{\hat D}\Phi
=\left({\rm det}\,{\hat D}\right)^{-1},\label{eqn:detDem}
\end{equation}
where the operator ${\hat D}$ is given by 
\begin{equation}
{\hat D}\equiv D_{\mu}D^{\mu}+m^2 \qquad {\rm and}
\qquad D_{\mu}\equiv\partial_{\mu}+iqA_{\mu}.\label{eqn:Dem}
\end{equation}

The determinant in equation~(\ref{eqn:detDem}) can be expressed as follows
\begin{equation}
\exp i \int d^4x\, {\cal L}_{corr}
= \left({\rm det}\,{\hat D}\right)^{-1}
=\exp -\,{\rm Tr}(\ln{\hat D})
=\exp -\int d^4x\, 
\langle t, {\bf x}\vert \,\ln {\hat D}\,\vert t, {\bf x}\rangle
\end{equation}
and in arriving at the last expression, following 
Schwinger~\cite{schwinger51}, we have chosen a complete and orthonormal
set of basis vectors $\vert t, {\bf x}\rangle$ to evaluate the trace 
of the operator $\ln {\hat D}$. 
>From the above equation it is easy to identify that
\begin{equation}
{\cal L}_{corr}= i\, \langle t, {\bf x}\vert\ln {\hat D}
\vert t, {\bf x}\rangle.
\end{equation}
Using the following integral representation for the operator 
$\ln {\hat D}$, 
\begin{equation}
\ln {\hat D}\equiv -\int_0^{\infty}{ds \over s}\, 
\exp -i({\hat D}-i\epsilon)s\label{eqn:lnD}
\end{equation}
(where $\epsilon \rightarrow 0^{+}$), the expression for ${\cal L}_{corr}$ 
can be written as 
\begin{equation}
{\cal L}_{corr}=-i\,\int_{0}^{\infty} {ds \over s}\,
e^{-i(m^2-i\epsilon)s}\,
K(t, {\bf x}, s\vert t, {\bf x}, 0),\label{eqn:lcorrDem}
\end{equation}
where 
\begin{equation}
K(t, {\bf x}, s\,\vert\, t, {\bf x}, 0)
=\langle t, {\bf x} \vert\, e^{-i{\hat H}s}\,\vert t, {\bf x}\rangle
\qquad {\rm and}\qquad
{\hat H}\equiv D_{\mu}D^{\mu}.\label{eqn:Hem}
\end{equation}
That is, $K(t, {\bf x}, s\,\vert\, t, {\bf x}, 0)$ is the kernel for a 
quantum mechanical particle in the coincidence limit (in four dimensions) 
described by the time evolution operator ${\hat H}$.
The variable $s$ that was introduced in~(\ref{eqn:lnD}) when the operator
$\ln{\hat D}$ was expressed in an integral form, acts as the time parameter 
for the quantum mechanical system.

The integral representation for the operator $\ln{\hat D}$ we have used 
above is divergent in the lower limit of the integral, {\it i.e.} near $s=0$.
This divergence should be regularized by subtracting from it another
divergent integral, {\it viz.} the integral representation of an operator
$\ln{\hat D_0}$, where ${\hat D_0} =(\partial^{\mu}\partial_{\mu}+m^2)$, 
the operator corresponding to that of a free quantum field. 
That is, to avoid the divergence, the integral representation for 
$\ln {\hat D}$ should actually be considered as
\begin{equation}
\ln {\hat D} -\ln {\hat D_0}\equiv 
-\int_0^{\infty}{ds \over s}\, 
\left\lbrace\exp-i({\hat D}-i\epsilon)s\;
-\;\exp-i({\hat D_0}-i\epsilon)s\right\rbrace.
\end{equation}
Or equivalently, the quantity ${\cal L}_{corr}^{0}$, which corresponds to 
the case of a free quantum field, can be subtracted from ${\cal L}_{corr}$ 
to obtain finite results.
The quantum mechanical kernel $K(t, {\bf x}, s\vert t, {\bf x}, 0)$ 
corresponding to the operator ${\hat D}_{0}$ is the kernel for a free 
particle in four dimensions,  {\it i.e.} $K(t, {\bf x}, s\vert t, 
{\bf x},0)=(1/16\pi^2 is^2)$. 
Substituting this quantity in the expression for ${\cal L}_{corr}$ above, 
we obtain that 
\begin{equation}
{\cal L}_{corr}^0= -\left({1\over{16 \pi^2}}\right)\,\int_{0}^{\infty} 
{ds \over s^3}\; e^{-i(m^2-i\epsilon)s}.\label{eqn:lcorrDem0}
\end{equation}
This is the expression which has to be subtracted from ${\cal L}_{corr}$
to yield a finite result. 
(It turns out that such a simple regularization scheme works for the cases 
we consider in this paper. 
In general, it may be necessary to use more complicated regularization 
schemes.)

\subsection{Evaluation of the effective lagrangian}

Now, consider a time independent electromagnetic background described by 
the vector potential
\begin{equation}
A^{\mu}=(\phi(x,y), 0, 0, \phi(x,y)),\label{eqn:A}
\end{equation}
where $\phi(x,y)$ is an arbitrary function of the coordinates $x$ and $y$.
The resulting electric field ${\bf E}$ and the magnetic field ${\bf B}$ are 
then given by
\begin{equation}
{\bf E}=-\left({\partial \phi \over {\partial x}}\,{\hat {\bf x}}
+{\partial \phi \over {\partial y}}\,{\hat {\bf y}}\right)\qquad;\qquad
{\bf B}=\left({\partial \phi \over {\partial y}}\,{\hat {\bf x}}
-{\partial \phi \over {\partial x}}\,{\hat {\bf y}}\right),\label{eqn:EB}
\end{equation}
where ${\hat {\bf x}}$ and ${\hat {\bf y}}$ are the unit vectors along the 
positive $x$ and $y$ axes repectively.
According to Maxwell's equations, in the absence of time dependence,  
the charge and the current densities, {\it viz.} $\rho$ and ${\bf j}$ 
that give rise to the above field configuration are
\begin{equation}
\rho = {\bf \nabla}.{\bf E}= -\left({\partial^2 \phi \over {\partial x^2}}
+ {\partial^2 \phi \over {\partial y^2}}\right)\qquad;\qquad
{\bf j} = {\bf \nabla}\times{\bf B}
= -\left({\partial^2 \phi \over {\partial x^2}}
+{\partial^2 \phi \over {\partial y^2}}\right)\,{\hat {\bf z}},
\label{eqn:rhoj}
\end{equation} 
where ${\hat {\bf z}}$ is the unit vector along the positive $z$ axis.
Therefore, if the functions $\rho$ and ${\bf j}$ are chosen such that they 
are finite and continuous everywhere and also vanish as $(x^2 + y^2)
\rightarrow\infty$, then the corresponding electric and magnetic fields 
given by equation~(\ref{eqn:EB}) will be confined to a finite extent 
in the $x-y$~plane.

It is obvious from equation~(\ref{eqn:EB}) that ${\cal G} = 2\,\left(
{\bf B}^2-{\bf E}^2\right) =0$ and ${\cal F}=-8\,\left({\bf E}.{\bf B}
\right)=0$ for this background field configuration.
(As an aside, note that this is an example of a field configuration 
other than that of a wave, for which ${\bf E}^2 -{\bf B}^2$ as well 
as ${\bf E}.{\bf B}$ are zero.) 
It is therefore a good candidate to test our conjecture.
The operator ${\hat H}$ that corresponds to the vector 
potential~(\ref{eqn:A}) is given by
\begin{equation}
{\hat H}\equiv \partial_t^2 - {\nabla}^2+2iq\phi (\partial_t
+ \partial_z).\label{eqn:HA}
\end{equation}
The kernel for the quantum mechanical particle described by the hamiltonian 
operator above can then be formally written as
\begin{equation}
K(t, {\bf x}, s\,\vert\, t,{\bf x}, 0) 
= \big\langle t, {\bf x} \big\vert 
\exp -i\left[\left({\partial_{t}}^2 - {\nabla}^2 
+2iq\phi (\partial_t + \partial_{z})\right)s\right]\,
\big\vert t, {\bf x}\big\rangle.
\end{equation}
Using the translational invariance of the hamiltonian operator 
${\hat H}$ along the time coordinate $t$ and the spatial coordinate 
$z$, we can express the above kernel as follows 
\begin{eqnarray}
K(t, {\bf x}, s\,\vert\, t, {\bf x}, 0) 
=\int_{-\infty}^{\infty} {d\omega \over 2\pi}
\int_{-\infty}^{\infty}{dp_z \over 2\pi}\; 
&& e^{i(\omega^2-p_z^2)s}\nonumber\\ 
&&\times\, \big\langle x, y\big\vert\, 
\exp -i\left[\left(-\partial_{x}^2\nonumber
-\partial_{y}^2 +2q(\omega - p_z)\phi\right)s\right]\, 
\big \vert x, y \big\rangle.
\end{eqnarray} 
Changing variables of integration in the expression above to 
$p_u=(p_z-\omega)/2$ and $p_v=(p_z+\omega)/2$, we find that
\begin{eqnarray}
K(t, {\bf x}, s\,\vert\, t, {\bf x}, 0) 
=\left({1 \over {2\pi^2}}\right)\;\int_{-\infty}^{\infty}dp_u\;
&& \int_{-\infty}^{\infty}dp_v\; e^{-4ip_u p_vs}\nonumber\\ 
&&\times\, \big\langle x, y\big\vert\, \exp -i\left[\left(-\partial_{x}^2
-\partial_{y}^2 -4q p_u \phi\right)s\right]\, \big \vert x, y \big\rangle.
\end{eqnarray} 
Performing the integrations over $p_v$ and the $p_u$ in that order, we 
obtain that
\begin{eqnarray}
K(t, {\bf x}, s\,\vert\, t, {\bf x}, 0) 
&=&\left({1 \over {\pi}}\right)\;\int_{-\infty}^{\infty}dp_u\; 
\delta(4p_us)\;  \big\langle x, y\big\vert\, 
\exp -i\left[\left(-\partial_{x}^2
-\partial_{y}^2 -4q p_u \phi\right)s\right]\, 
\big \vert x, y \big\rangle\nonumber\\
&=&\left({1 \over {4\pi s}}\right)\;
\big\langle x, y\big\vert\, \exp-i\left[\left(-\partial_{x}^2
-\partial_{y}^2\right)s\right]\, \big \vert x, y \big\rangle
=\left({1 \over {16 \pi^2 i s^2}}\right).\label{eqn:verify}
\end{eqnarray} 
(In arriving at the above result we have carried out the $p_v$ and the 
$p_u$ integrals first and then evaluated the matrix element. 
We show in Appendix~A that such an interchange of operations is 
valid by testing it in a specific example.) 
Substituting this expression for $K(t, {\bf x}, s\,\vert\, t, {\bf x}, 0)$
in~(\ref{eqn:lcorrDem}) we find that the resulting ${\cal L}_{corr}$ is 
the same as that of a free field. 
So, on regularization ${\cal L}_{corr}$ identically reduces to zero. 
This result then implies that in the time independent electromagnetic 
background we have considered here neither any particle production nor 
any vacuum polarization takes place.

As mentioned in the introduction, the effective lagrangian Schwinger 
had obtained for the constant electromagnetic background identically
vanishes when the gauge invariant quantities ${\cal G}$ and ${\cal F}$ 
are set to zero~\cite{schwinger51}.
Our result above agrees with Schwinger's result since a constant 
electromagnetic background would just correspond to choosing the 
function $\phi(x, y)$ above to be linear in the coordinates $x$ 
and/or $y$.
Having said that, we would like to stress here the following fact.
{\it In evaluating the effective lagrangian above we have not made 
any assumptions at all on the form of the function $\phi(x,y)$.}
Hence, our result above holds good for {\it any} time independent
electromagnetic background with vanishing ${\cal G}$ and ${\cal F}$.
Thus, in a way, our result here is more generic than Schwinger's result.

\section{Effective lagrangian for the example from gravity}\label{sec:grav}
\subsection{Preliminaries}

The system we shall consider in this section consists of a massive, real 
scalar field $\Phi$ coupled minimally to gravity.
It is described by the action
\begin{equation}
{\cal S}[g_{\mu\nu}, \Phi]
=\int d^4x\,\sqrt{-g}\;{\cal L}(g_{\mu\nu}, \Phi)
=\int d^4x\,\sqrt{-g}\;\left\lbrace {R \over {16 \pi}} 
+ {1 \over 2}\, g_{\mu\nu}\partial^{\mu}\Phi\partial^{\nu}\Phi 
- {1 \over 2}\, m^2 \Phi^2 \right\rbrace,\label{eqn:gravact}
\end{equation}
where $m$ is the mass of a single quantum of the scalar field and 
$g_{\mu \nu}$ is the metric tensor describing the gravitational 
background and we have set $G=1$ for convenience.
As it was done for the electromagnetic background in the last section, 
an effective lagrangian can be defined for the gravitational background 
as follows
\begin{equation}
\exp \,i\int d^4x\, \sqrt{-g}\; {\cal L}_{eff}(g_{\mu\nu})
\equiv\int{\cal D}\Phi\;
\exp \,i\, {\cal S}[\Phi, g_{\mu\nu}].
\end{equation} 
The effective lagrangian can then be expressed as 
${\cal L}_{eff} = {\cal L}_{grav} + {\cal L}_{corr}$, 
where ${\cal L}_{grav}=\left(R/16\pi\right)$, the 
lagrangian density for the gravitational background. 
Integrating the action for the scalar field in the above equation by parts 
and dropping the resulting surface terms, we find that ${\cal L}_{corr}$ 
can then be expressed as~\cite{bd82}
\begin{eqnarray}
\exp\,i\int d^4x\,\sqrt{-g}\; {\cal L}_{corr}(g_{\mu\nu})&=& 
\int{\cal D}\Phi\, \exp -i\int d^4x\,\sqrt{-g}\;
\left(\Phi {\hat D}\Phi\right)\nonumber\\
&=&\left({\rm det}\,{\hat D}\right)^{-1/2}
=\exp -{1 \over 2}\,{\rm Tr}(\ln{\hat D})\nonumber\\
&=&\exp -{1 \over 2}\,\int d^4x\,\sqrt{-g}\; 
\langle t, {\bf x}\vert \,\ln {\hat D}\,\vert t, 
{\bf x}\rangle,\label{eqn:detDgrav}
\end{eqnarray}
where the operator ${\hat D}$ is given by
\begin{equation}
{\hat D}\equiv {1 \over {\sqrt{-g}}}\partial_{\mu}
\left(g^{\mu\nu}\sqrt{-g}\partial_{\nu}\right) + m^2.\label{eqn:Dgrav}
\end{equation}
and, as it was done in the last section, we have introduced a complete set 
of orthonormal vectors $\vert t, {\bf x}\rangle$, to evaluate the trace.
>From equation~(\ref{eqn:detDgrav}) it is easy to identify that
${\cal L}_{corr}= (i/ 2)\, \langle t, {\bf x}\vert\ln {\hat D}
\vert t, {\bf x}\rangle$.
Using equation~(\ref{eqn:lnD}) ${\cal L}_{corr}$ above can then be 
written as
\begin{equation}
{\cal L}_{corr}= - {i \over 2}\, \int_{0}^{\infty}\, {ds \over s}\,
e^{-i(m^2-i\epsilon)s}\;
K(t, {\bf x}, s\vert t, {\bf  x}, 0),\label{eqn:lcorrDgrav}
\end{equation}
where
\begin{equation}
K(t, {\bf x}, s \vert t, {\bf x}, 0)=\langle t, {\bf x}\vert 
e^{-i{\hat H}s}\vert t, {\bf x}\rangle
\end{equation}
and the operator ${\hat H}$ is now given by
\begin{equation}
{\hat H}\equiv {1\over {\sqrt{-g}}}\partial_{\mu}\left(g^{\mu\nu}\sqrt{-g}
\partial_{\nu}\right).
\end{equation}
To obtain finite results, the quantity that has to be subtracted from 
${\cal L}_{corr}$, is then given by
\begin{equation}
{\cal L}_{corr}^{0}= -\left({1 \over 32 \pi^2}\right)\;
\int_{0}^{\infty}{ds \over s^3}\;e^{-i(m^2-i\epsilon)s},\label{eqn:lcorrDgrav0} 
\end{equation}
which corresponds to setting $g_{\mu\nu}=\eta_{\mu\nu}$ in the operator 
${\hat H}$ above.
(${\cal L}_{corr}^{0}$ given by equation~(\ref{eqn:lcorrDem0}) is twice
the ${\cal L}_{corr}^{0}$ above because the complex scalar field we had 
considered in the last section has twice the number of degrees of freedom 
as a real scalar field we are considering here.) 

\subsection{Evaluation of the effective lagrangian}

A gravitational background can be described by fourteen independent scalar
invariants constructed out of the Riemann curvature tensor~\cite{harvey90}. 
To verify our conjecture, we should evaluate ${\cal L}_{corr}$ defined 
in equation~(\ref{eqn:lcorrDgrav}) for a background for which all these
invariants vanish. 
And, of course, we need a background which is sufficiently simple for 
allowing the evaluation of ${\cal L}_{corr}$ in a closed form.

One such example is given by the spacetime described by the line element
\begin{equation}
ds^2 = (1+f(x,y))dt^2 - 2 f(x, y) dt dz - (1-f(x, y)) dz^2 - dx^2 - dy^2,
\label{eqn:met}
\end{equation} 
where $f(x, y)$ is an arbitrary function of the coordinates $x$ and $y$.
(This metric is a special case of the metric that appears 
in~\cite{koutras96}. 
It can be shown that all the fourteen algebraic invariants for this 
metric vanish identically~\cite{privcomn}.)
The non-zero components of the Ricci tensor for the above metric are
\begin{equation}
R^{00}=R^{33}=R^{30}=\left({1 \over 2}\right)\,
\left({\partial^2 f \over {\partial x^2}} 
+ {\partial^2 f \over {\partial y^2}}\right)
\end{equation}
and the Ricci scalar $R$ is zero.
Since the Ricci scalar $R$ is zero, the Einstein tensor is given by 
$G^{\mu\nu}=R^{\mu\nu}$ and the Einstein's equations reduce to 
$R^{\mu\nu}=8\pi\, T^{\mu\nu}$.
A pressureless steady flow of null dust with energy density $\rho=R^{00}$ 
traveling along the $z$-direction satisfies the above Einstein's equations 
and therefore gives rise to the metric~(\ref{eqn:met}).
Since $det(g_{\mu\nu})=-1$, the operator ${\hat H}$ corresponding to this 
metric is given by
\begin{equation}
{\hat H}= \partial_t^2 - \partial_z^2 - \partial_x^2 - \partial_y^2 
-f(\partial_t^2+\partial_z^2+2\partial_t \partial_z).
\end{equation}
Using the translational invariance along the $t$ and $z$ directions 
the kernel for the time evolution operator above can be written as
\begin{eqnarray}
K(t, {\bf x}, s\vert t, {\bf x}, 0)
= \int_{-\infty}^{\infty} {d\omega \over 2\pi}&& 
\int_{-\infty}^{\infty} {dp_z \over 2 \pi}\;
e^{i(\omega^2 - p_z^2)s}\nonumber\\
&&\times\,\langle x, y\vert \exp-i\left[\left(-\partial_x^2-\partial_y^2 
+ (\omega -p_z)^2\, f\right)s\right]\vert x, y\rangle.
\end{eqnarray}
Changing the variables of integration to $p_u=(p_z-\omega)/2$ and
$p_v=(p_z+\omega)/2$, we obtain that
\begin{eqnarray}
K(t, {\bf x}, s\vert t, {\bf x}, 0)
&=& \left({1 \over 2\pi^2}\right)\,
\int_{-\infty}^{\infty} dp_u \int_{-\infty}^{\infty} dp_v\; 
e^{-4ip_u p_vs}\nonumber\\
& &\qquad\qquad\qquad\quad\times\,
\langle x, y\vert \exp-i\left[\left(-\partial_x^2-\partial_y^2 + 4 
p_u^2\, f\right)s\right]\vert x, y\rangle\nonumber\\
&=&\left({1 \over \pi}\right)\,
\int_{-\infty}^{\infty} dp_u\; \delta(4p_us)\;
\langle x, y\vert \exp-i\left[\left(-\partial_x^2-\partial_y^2 + 4 
p_u^2\, f\right)s\right]\vert x, y\rangle\nonumber\\
&=& \left({1 \over 4 \pi s}\right)\,
\langle x, y\vert \exp-i\left[\left(-\partial_x^2-\partial_y^2\right)s\right]
\vert x, y\rangle= \left({1 \over {16 \pi^2 i s^2}}\right).
\end{eqnarray}
Substituting the above result in equation~(\ref{eqn:lcorrDgrav}) we find that
\begin{equation}
{\cal L}_{corr}= -\left({1 \over 32 \pi^2}\right)\,
\int_0^{\infty}{ds \over s}\, e^{-i(m^2-i\epsilon)s}.\label{eqn:lcorrfin}
\end{equation}
which on subtracting the quantity ${\cal L}_{corr}^{0}$ given by 
equation~(\ref{eqn:lcorrDgrav0}) reduces to zero.
This result again implies that in the gravitational background we have 
considered here neither any particle production nor any vacuum polarization 
takes place.
 
\section{Discussion}\label{sec:conclns}

The effective lagrangian provides a simple way of estimating the amount
of vacuum polarization and particle production in a classical background. 
For example, the background field is expected to induce vacuum instability 
and produce particles if and only if the effective lagrangian has an 
imaginary part. 
If the effective lagrangian vanishes for a particular background field, 
then no vacuum polarization or particle production takes place in such a 
field configuration.

In principle, this is an observable phenomenon since physical effects occur 
if the effective lagrangian happens to be non-zero.
For example, consider a constant electric field confined in space, say, 
the electric field between a pair of capacitor plates. 
In such a case, the imaginary part of effective lagrangian will be nonzero 
and the particle production will take place.
These particles that have been produced will get attracted towards the
capacitor plates thereby reducing the strength of the electric field 
between the plates. 
To maintain the original configuration intact, an external agency has to 
correct for this effect. 
We can therefore conclude that the above configuration---{\it viz.}, that 
of a constant electric field in a confined region---is not immune to 
quantum backreaction effects.
Such, physically observable, effects do occur even if the effective
lagrangian does not have an imaginary part.  
A typical example would be Casimir effect in flat spacetime.
It can be shown that for such a case the effective lagrangian is
non-zero and real; the real part, which depends on the separation
between the plates, can be related to the Casimir energy.
The resulting observable physical effect is the attraction between the 
Casimir plates. 
Left to themselves, the Casimir plates will move towards each other because 
of a force which is a quantum backreaction effect arising from the non-zero
real part of effective lagrangian. 
Once again, to maintain the original configuration---{\it viz.}, the original 
separation between the plates---an external agency has to correct for the
quantum backreaction effect.
 
In contrast to the above examples, backgrounds with vanishing effective 
lagrangian are `self-consistent' in the sense that no backreaction of the 
quantum field on the classical background occurs in these configurations.
This is a feature of certain backgrounds which, at least as far as the 
authors know, does not seem to have been noted in literature before. 
This aspect seems to be worthy of further study.

It should be possible to express the determinant of the operator 
${\hat D}$ (and hence the quantity ${\cal L}_{corr}$) appearing in
equations~(\ref{eqn:detDem}) and~(\ref{eqn:detDgrav}), at least 
formally, in terms of the invariant quantities describing the background.
In particular, one would expect the effective lagrangian to contain 
only those terms that are simple algebraic functions of the scalar 
invariants (otherwise renormalization would not be possible).
If so, the effective lagrangian would prove to be zero if all the 
invariants describing the background vanish identically.
Motivated by this fact, we put forward the conjecture that the regularized
${\cal L}_{corr}$ will prove to be zero for background field configurations 
for which all scalar invariants are zero. 
In other words, our conjecture implies that integrating out the degrees 
of freedom corresponding to the quantum field does not introduce any 
quantum corrections to the lagrangian describing classical backgrounds 
with vanishing scalar invariants.
 
We had also tested our conjecture with some specific examples.
For the electromagnetic background we have considered in 
section~\ref{sec:em} we had pointed out that the gauge invariant 
quantities ${\cal G}$ and ${\cal F}$ are zero and it can be easily
shown that quantities such as $\partial_\lambda F^{\mu\nu} 
\partial^{\lambda} F_{\mu\nu}$ and $\epsilon^{\lambda\rho\mu\nu} \partial_{\eta}F_{\lambda\rho}\partial^{\eta}F_{\mu\nu}$ also 
vanish identically. 
It is likely that all the gauge invariant quantities that can be 
constructed out of the vector potential~(\ref{eqn:A}) vanish 
identically.
For the gravitational example considered in section~\ref{sec:grav},
as mentioned before, it can be shown that all the fourteen algebraic 
invariants that can be constructed out of the Riemann tensor for the  
metric~(\ref{eqn:met}) vanish identically~\cite{privcomn}. 
Therefore, the vanishing of ${\cal L}_{corr}$ for these backgrounds 
is consistent with---and supports---our conjecture.

We would like to point out here the following fact.
The classical backgrounds we have presented in sections~\ref{sec:em}
and~\ref{sec:grav} are quite non-trivial though all the scalar 
invariants may vanish. 
They are not just flat space presented in an arbitrary gauge or a 
coordinate system.
The fact that a particle in these backgrounds will experience 
non-trivial forces acting on it ascertains this fact.

The examples that we had presented in sections~\ref{sec:em} 
and~\ref{sec:grav} are time independent examples.
As mentioned in the introduction, an example of a time dependent 
background for which the effective lagrangian proves to be zero 
is for that of a plane electromagnetic wave~\cite{schwinger51}.
(In appendix~B we rederive Schwinger's result using our technique.)
For the electromagnetic wave too it can be easily shown that apart from
the gauge invariant quantities ${\cal G}$ and ${\cal F}$, quantities such
as $\partial_{\lambda}F^{\mu\nu} \partial^{\lambda}F_{\mu\nu}$  and
$\epsilon^{\lambda\rho\mu\nu} \partial_{\eta}F_{\lambda\rho}
\partial^{\eta}F_{\mu\nu}$ also vanish identically.
It is, in fact, quite likely that all possible gauge invariant quantities 
vanish for the electromagnetic wave background thereby confirming our
conjecture.

Ideally, one would have liked to evaluate the effective lagrangian for an
arbitrary classical field configuration, vary the resulting effective 
lagrangian with respect to the classical fields and obtain the equations 
of motion for the classical background, thereby even taking into account  
the backreaction of the quantum field on the classical background. 
Since evaluating the effective lagrangian for an arbitrary classical 
background proves to be an impossible task, our approach to this entire 
problem has been a more practical one. 
The conjecture we have put forward in this paper is but the first step 
in this approach.
There exist deeper reasons in proposing this conjecture (with the 
danger of sounding obvious) and attempting to establish its validity 
with some specific examples. 
These  motivations are as follows.
The effective lagrangian may indeed prove to be zero for classical 
backgrounds for which all the scalar invariants are zero, but the 
converse need not be true.
That is, the effective lagrangian may prove to be zero even though 
some of the scalar invariants describing the background are non-zero.
Backgrounds with vanishing effective lagrangians but non-vanishing scalar 
invariants can help us identify the terms that will appear in the effective
lagrangian for the most general case.
Classifying such backgrounds will certainly prove to be a worthwhile 
exercise when evaluating the effective lagrangian for an arbitrary 
background is proving to be an impossible task.

\section*{Acknowledgments}
\noindent
LSK is being supported by the senior research fellowship of the
Council of Scientific and Industrial Research, India. 
RM wishes to thank Jawaharlal Nehru Centre for Advanced Scientific 
Research for the summer fellowship and IUCAA for hospitality.
The authors would also wish to thank Naresh Dadhich and Sukanya Sinha
for discussions.

\appendix
\section{}

In this appendix we illustrate the validity of carrying out the $p_v$ 
and the $p_u$ integrals first and then evaluating the matrix element 
in equation~(\ref{eqn:verify}) by testing it in a simple example.

Consider the case when $\phi(x,y)= x$.
This corresponds to a constant electromagnetic background with the
electric and magnetic fields given by ${\bf E}= -{\hat {\bf x}}$ and 
${\bf B} = -{\hat {\bf y}}$.
For this case, the operator ${\hat H}$ is given by
\begin{equation}
{\hat H} = \partial_t^2 - {\nabla}^2+2iq x (\partial_t + \partial_z). 
\end{equation}
The translational invariance of the above operator along the $t$, $y$ and 
$z$ directions can then be exploited to express the quantum mechanical 
kernel for the above operator as follows
\begin{eqnarray}
K(t, {\bf x}, s\,\vert\, t, {\bf x}, 0) 
=\int_{-\infty}^{\infty} {d\omega \over 2\pi}
\int_{-\infty}^{\infty} {dp_{z} \over 2\pi}&&\;
\int_{-\infty}^{\infty} {dp_y \over 2\pi}\;
e^{i(\omega^2-p_y^2-p_z^2)s}\nonumber\\
&&\times\,\big\langle x\big\vert\, \exp -i\left[\left(-d_{x}^2
+2q(\omega - p_z)x\right)s\right]\, 
\big \vert x \big\rangle.
\end{eqnarray} 
Carrying out the $p_y$ intergration and changing variables to 
$p_u=(p_z-\omega)/2$ and $p_v=(p_z+\omega)/2$, we obtain that
\begin{eqnarray}
K(t, {\bf x}, s\,\vert\, t, {\bf x}, 0) 
=\left({1 \over {2\pi^2 (4\pi is)^{1/2}}}\right)\;
\int_{-\infty}^{\infty}dp_v &&\;\int_{-\infty}^{\infty}dp_u\; 
e^{-4ip_u p_vs}\nonumber\\ 
&& \times\, \big\langle x\big\vert\, \exp -i\left[\left(-d_{x}^2
-4q p_u x\right)s\right]\, \big \vert x\big\rangle\label{eqn:kernel2}.
\end{eqnarray} 
The matrix element in the above equation corresponds to that of a 
quantum mechanical particle subjected to a constant force along the 
$x$-axis.
The matrix element above is then given by (see~\cite{dittrich94})
\begin{equation}
\big\langle x\big\vert\, \exp -i\left[\left(-d_{x}^2
-4q p_u x\right)s\right]\, \big \vert x\big\rangle
= \left({1 \over {(4\pi is)^{1/2}}}\right)\;
\exp -4i\left(q p_u xs + {1 \over 3} q^2 p_u^2 s^3\right).
\end{equation}
Substituting this expression in the kernel~(\ref{eqn:kernel2}), 
we obtain that
\begin{eqnarray}
K(t, {\bf x}, s\,\vert\, t, {\bf x}, 0) 
&=&\left({1 \over {8\pi^3 is}}\right)\;
\int_{-\infty}^{\infty}dp_v\;\int_{-\infty}^{\infty}dp_u\; 
\exp -4i\left({1 \over 3} q^2 p_u^2 s^3 +p_u s(qx+p_v)\right)\nonumber\\
&=&\left({1 \over {8\pi^3 is}}\right)\;
\left({3\pi \over {4iq^2s^3}}\right)^{1/2}\;
\int_{-\infty}^{\infty}dp_v\; 
\exp \left({3i \over q^2s}(p_v + qx)^2\right)\nonumber\\
&=& \left({1 \over 16 \pi^2 i s^2}\right),
\end{eqnarray}
which is the result quoted in the text.

\section{}

In this appendix we rederive Schwinger's result for the electromagnetic 
wave background using our technique.
The plane electromagnetic wave can be described by the vector potential
\begin{equation}
A_{\mu}=(0, 1, 0, 0)f(t-z)
\end{equation}
where $f(t-z)$ is an arbitrary function of $(t-z)$.
The operator ${\hat H}$ corresponding to this vector potential is then
given by
\begin{equation}
{\hat H}= \partial_t^2 -\partial_x^2 -\partial_y^2-\partial_z^2
+2iqf\partial_x +q^2f^2
\end{equation}
and in terms of the null coordinates $u=(t-z)$ and $v=(t+z)$ the above 
operator reduces to
\begin{equation}
{\hat H} = 4\partial_u\partial_v -\partial_x^2 -\partial_y^2
+2iqf(u)\partial_x +q^2f^2(u).
\end{equation}
The corresponding quantum mechanical kernel can then be formally 
expressed as
\begin{eqnarray}
&& K(u, x, y, v, s\vert u, x, y, v, 0)\nonumber\\
&&\qquad\qquad\qquad\quad
=\langle u, x, y, v \vert \exp -i\left[\left(4\partial_u\partial_v 
-\partial_x^2 -\partial_y^2+2iqf\partial_x +q^2f^2\right)s\right]
\vert u, x, y, v \rangle.
\end{eqnarray}
Exploiting the translational invariance of the operator ${\hat H}$ 
along the $x$, $y$ and the $v$ coordinates we can write the above kernel
as
\begin{eqnarray}
K(u, x, y, v, s\vert u, x, y, v, 0)
=\int_{-\infty}^{\infty}{dp_x \over 2\pi}&& 
\int_{-\infty}^{\infty}{dp_y \over 2\pi}\;
\int_{-\infty}^{\infty}{dp_v \over 2 \pi}\;
e^{-i(p_x^2+p_y^2)s}\nonumber\\
&&\times\;2\;\langle u \vert \exp -i\left[\left(-4ip_v d_u
-2qp_x f +q^2f^2\right)s\right]\vert u \rangle,\label{eqn:kuv}
\end{eqnarray}
where the factor $2$ is the Jacobian of the transformation between the 
conjugate momenta $(\omega, p_x)$ and $(p_u, p_v)$ corresponding to 
the coordinates $(t, z)$ and $(u, v)$ respectively.

The matrix element in the above equation corresponds to the quantum 
mechanical kernel for a time evolution operator given by
\begin{equation}
{\hat H}_1 = -4ip_v d_u -2qp_xf(u) + q^2 f^2(u).
\end{equation}
The normalized solution $\psi_E(u)$ to the time independent 
Schr{\" o}dinger equation for the operator ${\hat H}_1$ corresponding 
to an energy eigen value $E$ is then given by
\begin{equation}
\psi_E(u)= \left({1\over 8\pi p_v}\right)^{1/2}\;
e^{iqE/4 p_v} \exp -i \left(h(u)/4 p_v\right),
\end{equation}
where
\begin{equation}
h(u)=-\int du \left(2qp_xf(u) - q^2 f^2(u)\right).
\end{equation}
The matrix element can now be evaluated with the help of the 
Feynman-Kac formula~\cite{fandh65} as follows
\begin{eqnarray}
\langle u \vert e^{-i{\hat H}_1 s}\vert u' \rangle 
&=&\int_{-\infty}^{\infty} dE\; \psi_E(u)\; 
\psi_E^*(u')\;e^{-iEs}\nonumber\\  
&=&\left({1 \over 8\pi p_v}\right)\; 
\exp-i\bigg\lbrace(h(u) - h(u'))/4p_v\bigg\rbrace\; 
\int_{-\infty}^{\infty} dE\; \exp i\left(E(u-u')/4 p_v\right)\;
e^{-iEs}\nonumber\\
&=& \exp- i\bigg\lbrace(h(u) - h(u'))/4p_v\bigg\rbrace\;
\delta\left(u-u'-4 p_v s \right) 
\end{eqnarray} 
and in the coincidence limit $u=u'$, the matrix element reduces to a 
Dirac delta function {\it i.e.}
\begin{equation}
\langle u \vert \exp -i{\hat H}_1 s\vert u \rangle =\delta(4 p_v s).
\end{equation} 
Substituting this result in equation~(\ref{eqn:kuv}), we obtain that
\begin{eqnarray}
K(u, x, y, v, s\vert u, x, y, v, 0)
&=&\left({2 \over {(4\pi i s)^{1/2}}}\right)\;
\int_{-\infty}^{\infty}{dp_x \over 2\pi}\; e^{-ip_x^2s} 
\int_{-\infty}^{\infty}{dp_v \over 2 \pi}\; \delta(4p_v s)\nonumber\\
&=&\left({2 \over {4\pi i s}}\right)\;
\int_{-\infty}^{\infty}{dp_v \over 2 \pi} \left({1 \over4s}\right)\;
\delta(p_v)=\left({1 \over {16 \pi^2 i s^2}}\right).
\end{eqnarray} 
When the above kernel is substituted in equation~(\ref{eqn:lcorrDem}) 
we find that the resulting ${\cal L}_{corr}$ is the same as that of 
${\cal L}_{corr}^0$, which on regularization reduces identically 
to zero.

\end{document}